# ChatGPT's advice drives moral judgments with or without justification

Sebastian Krügel[1]    Andreas Ostermaier[2]    Matthias Uhl[1, ‡]

**Abstract**

Why do users follow moral advice from chatbots? A chatbot is not an authoritative moral advisor, but it can generate seemingly plausible arguments. Users do not follow reasoned more readily than unreasoned advice, though, we find in an experiment. However, this is also true if we attribute advice to a moral advisor, not a chatbot. Hence, it seems that advice offers users a cheap way to escape from a moral dilemma. This is a concern that chatbots do not raise, but they exacerbate it as they make advice easily accessible. We conclude that it takes ethical in addition to digital literacy to harness users against moral advice from chatbots.

[1] Faculty of Business, Economics and Social Sciences, University of Hohenheim, Schloss Hohenheim 1, 70599 Stuttgart, Germany.
[2] Institute for a Sustainable Economy, Zeppelin University, Am Seemoser Horn 20, 88045 Friedrichshafen, Germany.
‡ matthias.uhl@uni-hohenheim.de.

**ChatGPT's advice drives moral judgments with or without justification**

**Introduction**

Chatbots like OpenAI's ChatGPT or Google's Bard have made AI accessible to common internet users. While AI has conquered all realms of life, chatbots enable laypeople to use AI to solve tasks that many believed are reserved to human reason. In particular, users tend to follow advice from chatbots and rely on them to make their own decisions. This is also true for moral judgments, recent research shows, although nothing commends a "stochastic parrot" without any moral convictions as a moral advisor (Krügel et al., 2023b). Advice-taking from chatbots can have dramatic consequences, as in the case of a man who ended his life for the benefit of our planet, following a weeks-long conversation with a chatbot (El Atillah, 2023).

The seemingly blind trust of users into chatbots is surprising in the light of prior research on the role of advice in human decision-making. The literature documents that we are reluctant to follow advice from others for a number of reasons. These relate, along with the advisee, to the advisor, the advice, and the task (Bonaccio & Dalal, 2006; Kämmer et al., 2023). We prefer advisors who are similar to us and likely reason in a similar way to advise us on matters of judgment. Rather than plain recommendations, we would rather have advice that includes arguments. Moreover, the moral domain seems more than anything else to be reserved to human reasoning. There is little to suggest that chatbots can influence our moral judgment. Nonetheless, evidence that the influence of moral advice by artificial advisors is large—about as large as of human advice indeed—is accumulating (Krügel et al., 2022, 2023a, 2023b).

A possible reason why moral advice by chatbots is so influential is their capability to argue. Although a chatbot's advice is eventually random and arbitrarily tells users to make one decision as much as another in a moral dilemma (Krügel et al., 2023b), it is in either case supported by some argument. Of course, a chatbot cannot reason; it recombines words in a random but sophisticated way into an argument. Nonetheless, this argument may contain reasons that convince users to follow, or it may at least make the chatbot's advice look sounder. That said, the nature of the moral problem suggests an alternative explanation. Moral dilemmas impose a burden on decision-makers, who are then willing to jump at any advice for relief (Kämmer et al., 2023). We examine whether the argument made by a chatbot in support of its advice is what makes the difference.

We conducted an online experiment to answer this question. Our subjects were presented with a version of the trolley dilemma, which is about whether to sacrifice one life to save five (Foot, 1967; Greene et al., 2001; Thomson, 1976). They were asked to judge what is the right thing to do in this case, but first take advice. The advice came either with or without an argument to justify the recommendation, and we either attributed it to a moral advisor or revealed ChatGPT as the advisor. We used advice both for and against sacrificing one life to save five. Our results confirm that moral advice by ChatGPT influences users' judgment. However, the influence of advice does not statistically differ depending on whether it is reasoned or unreasoned. Interestingly, this also holds if it is attributed to a moral advisor. We conclude that it is the task—the moral dilemma—, not chatbots' capability to argue that explains their influence on users' moral judgments.

This finding has important implications. Why users' susceptibility to moral advice does not arise from chatbots apparently, chatbots can readily offer the advice users desire and, thus, influence their judgements. Hence, chatbots give developers considerable power over our moral judgments. Well-meaning developers can train chatbots to decline to give moral advice. Even then, however, chatbots may fail to recognize moral problems. Therefore, the question is how we can harness users against falling for chatbots' advice. We argue that, along with digital literacy, it takes ethical literacy. We believe that users who know about the limitations of chatbots and who have firm moral convictions will not that easily rely on "stochastic parrots" to advise them.

Our paper proceeds in three steps. Next, we review the literature on the use of advice, and particularly moral advice. We then present our methods and summarize our results. Finally, we conclude with a discussion of our findings.

**Theory**

Why do advisees follow or not follow advice? This is a key question in research on advice-based decision-making or judge–advisor systems (Bonaccio & Dalal, 2006; Kämmer et al., 2023). The literature shows that advisees tend to discount advice in terms of how much weight they give to the advisor's relative to their own judgment. For example, although advisees revise estimates after receiving advice on estimation tasks, they stick (too much) to their own initial estimates (Bonaccio & Dalal, 2006; Yaniv, 2004). Recent research extends the notion of advisor to include algorithms, chatbots, and the like, with a focus on how much advisees follow advice from such artificial relative to human advisors (Hütter & Fiedler, 2019; Logg et al., 2019; Krügel et al., 2022, 2023a, 2023b; Leib et al., 2024). Utilization of advice depends on a number of factors, which can be attributed to the advisee, the advisor, the task, and the advice (Bonaccio & Dalal, 2006; Kämmer et al., 2023).

*Advisor*

Whether advice is followed depends on characteristics of the advisor, particularly her or his expertise and trustworthiness. For example, we infer an advisor's expertise from her or his track record, experience, or even confidence (Bonaccio & Dalal, 2006; Kämmer et al., 2023). When it comes to matters of judgment or taste, however, advisees do not necessarily want advisors to be impartial experts, but to be similar to them (Kämmer et al., 2023). For example, we prefer advice on songs, movies, or restaurants from friends rather than strangers (Yaniv et al., 2011; Eggleston et al., 2015; Tuk et al., 2019). An explanation for why we discount advice from others relative to our own judgment is that we do not know about their reasoning as much as we know about ours. Assuming that similar others reason similarly to us, we think that similarity reduces this asymmetry in reasoning, and we are more willing to follow advice from them (Kämmer et al., 2023).

Chatbots are human-like as a technology, but they are less similar to us than any stranger. Moreover, they cannot reason, in the first place, so their "reasoning" cannot be similar to ours. Instead, they combine words in a likely but random way to imitate human speech. Thus, advisees should be reluctant to follow a chatbot's advice. That said, prior evidence is mixed on whether advisees are less or more susceptible to advice from algorithms, some finding algorithm aversion (Dietvorst et al., 2015), some appreciation (Logg et al., 2019), some neither (Hütter & Fiedler, 2019; Krügel et al., 2022, 2023a, 2023b; Leib et al., 2024). Recently, evidence has accumulated, first, that advisees utilize advice from artificial advisors, such as algorithms or chatbots, and second, that utilization does not depend on whether advisors are artificial or human. Although the similarity argument suggests otherwise, we should therefore expect to see decision-makers follow advice from a chatbot, even if it is revealed as such.

*Task*

That said, the tasks we are advised on are heterogeneous, and we are more willing to accept advice on some than on others. The literature has focused on small set of tasks, including estimation, forecasting, gambling, and math tasks (Bonaccio & Dalal, 2006; Kämmer et al., 2023). Research on advice-taking on matters of judgment or taste are rare (Yaniv et al., 2011;

Tzioti et al., 2014; Eggleston et al., 2015; Logg et al., 2019; Tuk et al., 2019). Even within this small set of tasks, utilization of advice varies. For example, advisees are more reluctant to follow advice on estimation than choice tasks. There is also evidence that advisees seek (but not that they utilize) advice more readily if their task is difficult (Kämmer et al., 2023). Taken together, these findings support the intuition that the task is important for whether advice is utilized. However, evidence on matters of judgment, let alone moral judgement rather than taste, is limited.

The available evidence shows that advice is readily taken in the moral domain. This is unsurprising because moral dilemmas are difficult problems, which cannot be (re)solved by definition. Hence, we have any reason to expect that decision-makers utilize moral advice if they are offered it (Kämmer et al., 2023). In turn, it is surprising that decision-makers do not mind advice coming from algorithms (Hütter & Fiedler, 2019; Krügel et al., 2022, 2023a) or chatbots (Krügel et al., 2023b; Leib et al., 2024). The moral domain, more than anything, seems to be reserved to human reasoning. Machines cannot reason, though, and they do not have any moral convictions. They may advice for on solution as much as for the other. Their advice is eventually random, and users who follow it base a difficult moral decision on the toss of a potentially unfair coin. That said, prior evidence shows that users follow moral advice, also if is revealed as coming from a chatbot (Krügel et al., 2023b).

*Advice*

Advice varies in quality, and advisees want to follow good but not bad advice. The quality of advice is hard to establish for advisees, though. This is why we tend to infer it from the quality of the advisor, which we often find easier to observe (Bailey et al., 2023). However, there are attributes to advice that potentially allow advisees to assess its quality, independently of the advisor. In particular, advice can come with or without arguments to support it. If advice is supported by arguments, advisees can see that it is reasoned, verify whether they agree with the reasons, and decide whether they consequently utilize it. There is evidence that analytical advice is more readily followed than intuitive advice (e.g., "market research says …" versus "my intuition says …"). This effect is not independent of other factors, though. For example, intuitive advice is followed more if the task is a matter of taste (Ribeiro et al., 2020; Tzioti et al., 2014).

The capability to argue is a fascinating feature of chatbots. If asked for advice, they can, in addition to telling us what to do, argue why. As noted above, chatbots cannot actually reason like humans. They are "stochastic parrots," which recombine words in a sophisticated but eventually random way (Bender et al., 2021). Therefore, advisees might, on the one hand, ignore arguments forged by them as babbling. On the other hand, however, these arguments can be, or at least sound, reasonable, and they can make advice look more reasoned. Moreover, it is not obvious why users would ignore chatbots' arguments, but not discount their advice, whether it comes with (Krügel et al., 2023b; Leib et al., 2024) or without arguments (Hütter & Fiedler, 2019; Krügel et al., 2022, 2023a). Taken together, there are both reasons to expect that users consider chatbots' arguments and that they ignore them. Hence, whether their arguments drive chatbots' influence on advisees is an open question.

The question of how much arguments drive the utilization of advice arises particularly in the moral domain. Research shows, we noted, that we do follow moral advice, both with and without arguments to support it. However, do we utilize advice more readily if it is reasoned rather than unreasoned, and particularly reasoned advice from a chatbot? On the one hand, arguments likely increase the perceived quality of advice, and they may indeed contain cues that leads advisees to change their mind, who would not follow plain recommendations. This may also be true for advice from a chatbot, although it creates its arguments randomly. On the other hand, moral dilemmas are ambiguous problems, and advisees might discount the arguments along with the advice as intuition-based babbling. Thus, decision-makers might well follow moral advice but disregard the arguments, and whether there are arguments. In particular, it remains to be seen whether advisees put more or less weight on arguments from a chatbot than from a human advisor.

**Methods**

We conducted our online experiment in May 2023. The subjects were recruited from CloudResearch's Prime Panels (Litman et al., 2017). Online platforms such as CloudResearch or MTurk produce reliable and valid results for a variety of tasks and measurements (Goodman et al., 2013; Hauser & Schwarz, 2016; Chandler et al., 2019). Prime Panels participants must pass standard screening questions to participate in a study, and they are more diverse and representative of the US population than MTurk participants (Chandler et al., 2019). Participation took about 5 minutes and paid $1.25.

Subjects faced the "bridge" version of the trolley dilemma. This dilemma asks whether it is right to push a stranger from a bridge onto the tracks of a tram to block a run-away trolley, which will otherwise kill five people down the tracks. The stranger will die if he is pushed down (Foot, 1967; Greene et al., 2001; Thomson, 1976). We chose the bridge version because the influence of moral advice on subjects' judgments on this dilemma turned out surprisingly strong in prior research (Krügel et al., 2023b), which gives us room to investigate how much an argument in support of the advice explains this influence. Different ethical schools suggest different judgments in the bridge dilemma. Utilitarians, who strive for the greatest good of the greatest number, would rather sacrifice one life for five. Kantianism, on the contrary, argues against using a fellow human as a mere means to another end—a stopper, in this case. Empirically, most people judge that the stranger should not be pushed (Awad et al., 2020; Plunkett & Greene, 2019). Our focus, however, is not on subjects' judgments, but on how advice from a chatbot influences those judgments.

Before our subjects made their own judgments, they read a transcript of a conversation with ChatGPT, consisting of a question and ChatGPT's answer, which we borrowed from prior research (Krügel et al., 2023b). ChatGPT was asked several questions such as "Would it be right to sacrifice one person to save five others?" These questions were the same except for wording, but ChatGPT advised sometimes for, sometimes against sacrificing one life to save five. The answers were obtained from ChatGPT in December 2022, soon after its release (Krügel et al., 2023b). Note that the purpose of this study is not to test the latest version of ChatGPT. In fact, ChatGPT now recognizes the trolley dilemma reliably and no longer advises either for or against sacrificing one life to save five. Instead, our purpose is to examine whether the influence of moral advice by chatbots, like ChatGPT, on human judgment is driven by their capability to argue. This capability is distinctive of LLMs and it is reasonable to assume that it leverages their influence (Krügel et al., 2023b).

We adopted two out of the six transcripts of chats from that study, one where ChatGPT argues for and one where it argues against sacrificing one life. We chose these two versions because they each contained one sentence with clear advice as well as several other sentences arguing for this advice. Figure 1 (top) depicts the latter for illustration. For our experiment, we modified ChatGPT's answers in two ways. First, to examine the incremental effect of the argument, we created modified versions of the two transcripts without ChatGPT's argument. That is, we did not change the wording of ChatGPT's advice, but we removed the sentences containing the argument for the advice. Figure 1 (bottom) shows the modified version of the transcript in Figure 1 (top). Second, to study the differential effect of the source of advice (ChatGPT versus a moral advisor), we removed anything that might hint to ChatGPT, both from the original transcript with argument and the modified version without argument.

—Insert Figure 1 about here.—

Overall, the experiment featured 8 (= 2 × 2 × 2) conditions. The answer in the transcript advised either for or against sacrificing one life to save five, was either reasoned or unreasoned, and was attributed to either ChatGPT or a moral advisor. In the former case, we followed Krügel et al. (2023b) and introduced ChatGPT as "an AI-powered chatbot, which uses deep learning to talk like a human." In the latter case, any reference to ChatGPT was suppressed.

The experiment received an ethical review by the German Association for Experimental Economic Research (https://gfew.de/en) and was approved without restriction. We conducted our study according to the principles of the Declaration of Helsinki. Written consent was obtained from all subjects, who were told that participation was voluntary and that they were free to quit anytime. The study was preregistered at AsPredicted.org (https://aspredicted.org/9S3_M9W). Screenshots of the questionnaire are attached as Supplementary Information.

We needed our subjects to understand (a) who or what advised them and (b) how they advised them to study the effect of these two factors on their moral judgment. Therefore, we included two post-experimental multiple-choice questions, which asked them to identify their advisor (ChatGPT or a moral advisor) and advice (for or against sacrificing one life). As preregistered, we only consider the responses of the subjects who answered both of these questions correctly. In total, 1,269 subjects passed the comprehension questions and their age averages 44.5 years, ranging from 18 to 86. 60% are female; 39%, male. 1% are non-binary or did not indicate their gender. Table 1 summarizes information about our subjects, broken down by advisor (ChatGPT versus moral advisor) and advice (with versus without argument). The experimental conditions do not statistically differ in terms of subjects' age and gender. Subjects' self-assessment of how ethical they consider themselves relative to other subjects was also similar across conditions. Like in related research, subjects considered themselves more ethical than others (e.g., Krügel et al. 2022).

—Insert Table 1 about here.—

## Results

### *Influence of reasoned vs. unreasoned advice on moral judgment*

Figure 2 summarizes our subjects' judgments, depending on whether their advice came with or without an argument and whether their source of advice was ChatGPT or a moral advisor. We note, first, that the effects of advice with argument closely replicate those from prior research. In Krügel et al. (2023b), 72 percent of the subjects think it is right to push the stranger off the bridge to save the five people on the track if a moral advisor advises them so. If the advisor advises against pushing the stranger, only 17 percent think so. If ChatGPT is revealed as the advisor, 63 or 25 percent of the subjects of that study think it is right to push the stranger off, depending on the advice. In our study, 75 percent of the subjects think it is right if a moral advisor advises them so, with an argument in support, but only 18 percent think so if advised against pushing him (75 vs. 18 percent, $\chi^2(1, 334) = 104.34$, $p < 0.01$). Likewise, if ChatGPT is revealed as the advisor and advice is supported by an argument, 70 or 27 percent think so, depending on the advice (70 vs. 27 percent, $\chi^2(1, 332) = 59.09$, $p < 0.01$). Hence, as in Krügel et al. (2023b), moral advice that comes with an argument strongly influences subjects' moral judgment.

Second, Figure 2 shows that the percentages hardly change if advice is given without an argument. If advised so, 78 percent of the subjects think it is right to push the stranger off. These percentages do not statistically differ from the case where advice comes with an argument (75 vs. 78 percent, $\chi^2(1, 335) = 0.30$, $p = 0.58$). If advised against pushing him, 19 percent think so, which again does not differ from the case with argument (18 vs. 19 percent, $\chi^2(1, 287) < 0.01$, $p > 0.99$). In the case where ChatGPT is revealed as the advisor, 77 percent think it is right to push the stranger off if advised so (70 vs. 77 percent, $\chi^2(1, 346) = 1.71$, $p = 0.19$), but only 23 percent, if advised against (27 vs. 23 percent, $\chi^2(1, 301) = 0.48$, $p = 0.49$). In fact, the percentages do not differ depending on whether the advice comes from a moral advisor or ChatGPT or whether the advice comes with or without a justification ($\chi^2(3, 681) = 3.21$, $p = 0.36$ if the advice is to push the stranger off; $\chi^2(3, 588) = 4.71$, $p = 0.19$ otherwise). Hence, reasoned advice does not influence judgements any more than unreasoned advice.

—Insert Figure 2 about here.—

*Potential psychological mechanism*

After the subjects submitted their moral judgments, we asked them to assess how plausible they found the advice, and how authoritative they found their advisor, both on a scale ranging from 0 to 6. Figure 3 illustrates the assessment of the moral authority of the advisor and of the plausibility of the advice by those subjects who followed the advice, depending on whether the source was ChatGPT or a moral advisor and on whether the advice came with or without an argument. The subjects who were told that they received advice from ChatGPT rated the moral authority of the source of advice significantly lower than those who were told that the advice came from a moral advisor (2.79 vs. 3.19, $W = 34{,}488$, $p = 0.023$ with argument; 2.29 vs. 2.83, $W = 31{,}815$, $p = 0.003$ without argument). At the same time, these subjects rated the plausibility of the advice from ChatGPT significantly higher than subjects who received advice from a moral advisor (4.31 vs. 3.79, $W = 24{,}855$, $p < 0.001$ with argument; 3.49 vs. 3.13, $W = 24{,}380$, $p = 0.03$ without argument). However, the advice was identical in both cases, and advice by ChatGPT could not possibly be more plausible.

—Insert Figure 3 about here.—

A psychological mechanism that is consistent with our data is an ex-post rationalization of moral judgments (Haidt, 2001). Ethical dilemmas represent difficult-to-solve tasks in which a choice must be made between two "bads." Such tasks generate emotional and cognitive distress for decision-makers. Advice on how to solve moral dilemmas is welcome, regardless of whether it is well-founded or not. It helps reduce psychological costs and is therefore readily accepted. To justify adherence to the advice to themselves, advisees refer to the moral authority of the advisor or the plausibility of the advice. The participants in our experiment were aware that ChatGPT has no particular moral authority. To compensate, they assessed the plausibility of ChatGPT's advice to be higher. That is, the increased plausibility rating substitutes for the lack of moral authority of ChatGPT as a moral advisor.

It is, of course, possible that the causality is reversed. Subjects who were told that they were advised by ChatGPT did find the advice more plausible and, consequently, followed it more often, despite ChatGPT's lack of moral authority. This explanation is unlikely, though. It assumes that randomization failed, such that subjects who were told that their advisor was ChatGPT happened to find the advice more plausible than those who were not, with a sample size of several hundred subjects. Moreover, subjects found advice with an argument more plausible than advice without (3.95 vs. 3.29, $W = 41{,}937$, $p < 0.001$ for ChatGPT; 3.76 vs. 3.12, $W = 38{,}726$, $p < 0.001$, for the moral advisor), and they considered ChatGPT less authoritative than a moral advisor (2.28 vs. 2.91, $W = 238{,}396$, $p < 0.001$). Nonetheless, advice with argument does not influence judgements any more than advice without argument, and ChatGPT's influence is about the same as that of a moral advisor. The influence of advice is not apparently driven by its plausibility or by the moral authority of its source. Otherwise, we should see a stronger effect of advice with an argument on subjects' judgments than of advice without, and a weaker effect of advice from ChatGPT than from a moral advisor, following subjects' assessments of plausibility and moral authority. The data are more in line with an ex-post rationalization of moral judgements.

## Discussion

The purpose of this study is to understand why users follow moral advice from chatbots so readily. A chatbot qualifies hardly as an authoritative moral advisor, but it can argue, and its arguments might convince users to follow its advice. Hence, we conjectured that users follow chatbots' advice because these argue for it and provide reasons rather than make plain recommendations. However, that is not what we find. Users do not follow reasoned advice from ChatGPT any more than unreasoned advice, although they do find the former more plausible. Interestingly, we observe the same if advice is not attributed to ChatGPT but to a moral advisor. Why do users then follow advice from chatbots, if it is not for chatbots' arguments? Apparently, we seek cheap ways out of moral dilemmas by following any advice we get, reasoned or unreasoned, and regardless of whether the advisor is human or a bot. Rather than as a source of arguments, advice eases our burden.

Moreover, we also find evidence that chatbot users, rather than follow advice because they find it plausible, find it plausible because they followed it, regardless of whether it is reasoned or unreasoned. This perceived plausibility is a case of ex-post-facto rationalization (Haidt, 2001). It turns out that identical advice is considered more or less plausible by advisees, depending on whether it is attributed to a chatbot rather than a moral advisor. At the same time, a chatbot is considered less morally authoritative than a moral advisor. Thus, the perceived plausibility of the argument is really driven by the advisor. Users who follow a chatbot's advice delude themselves into believing that they bow to the force of the argument, while they grant a "stochastic parrot" moral authority over themselves indeed. Users' factual, ex-post rationalized adoption of chatbots' advice gives developers substantial manipulative power over users. It is, therefore, a crucial task for future research to shed more light on these psychological mechanisms.

The spread of AI raises questions about how users can be empowered to use it responsibly. In particular, chatbots make AI accessible to lay users, who fail to understand or consider its limitations. Thus, there is growing evidence that users follow advice from AI-powered chatbots for their moral judgments and decisions, although these do not lend themselves as moral advisors, even if the source of the advice is perfectly transparent (Krügel et al., 2022, 2023b). Our findings suggest that chatbots' influence on users does not depend on their capability to argue, but to just give advice. In moral dilemmas, we seek guidance and gratefully accept others' stances. If people engage in ex-post rationalization, it matters little where the advice comes from and how plausible it is. Being susceptible to moral influence is not a problem that arises because of AI, but the ubiquity of AI aggravates its social implications. In the future, our AI advisor will be our closest companion, which we carry in our pocket.

How can we mitigate this concern? On the one hand, chatbots can be trained to decline to advise users on moral dilemmas. However, chatbots may fail to recognize these, and moral dilemmas are baked into many everyday decisions. Chatbots might often need to decline, which makes them fairly useless. On the other hand, we users can be trained to challenge rather than naively trust moral advice of chatbots. Thus, we will need to promote digital literacy. A basic understanding of how chatbots work ("stochastic parrots") may be enough to foster critical and responsible use of AI-companions. That said, our results also show that the problem is the lack of an ethical compass at the individual level. Many of us do not have a solid ethical stance, and this makes us so accessible to AI-powered advice in the first place. In addition to digital literacy, it is therefore necessary to improve ethical literacy. If the latter is achieved, a critical approach to AI will come naturally.

**Competing Interests**

The authors declare no competing interests.

**Data Availability**

The data will be made available upon request by the corresponding author of this publication.

**Ethical Approval**

This study was performed in line with the principles of the Declaration of Helsinki. Approval was granted by the German Association for Experimental Economic Research.

**Informed Consent**

Informed consent was obtained from all participants.

**Author Contributions**

S.K., A.O., and M.U. designed and performed the study, analyzed the data, and wrote the report together.

**Funding**

This work was supported by Bavarian Research Institute for Digital Transformation.

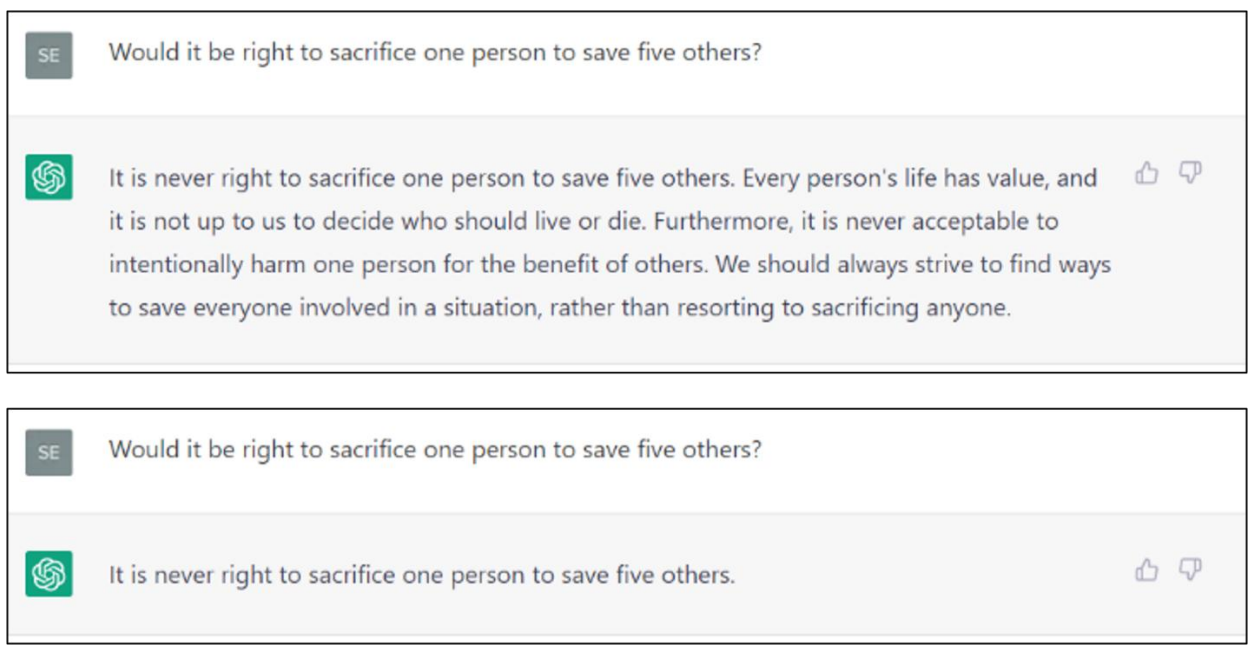

**Figure. 1.** Advice by ChatGPT against sacrificing one life to save five with an argument (top) and without (bottom).

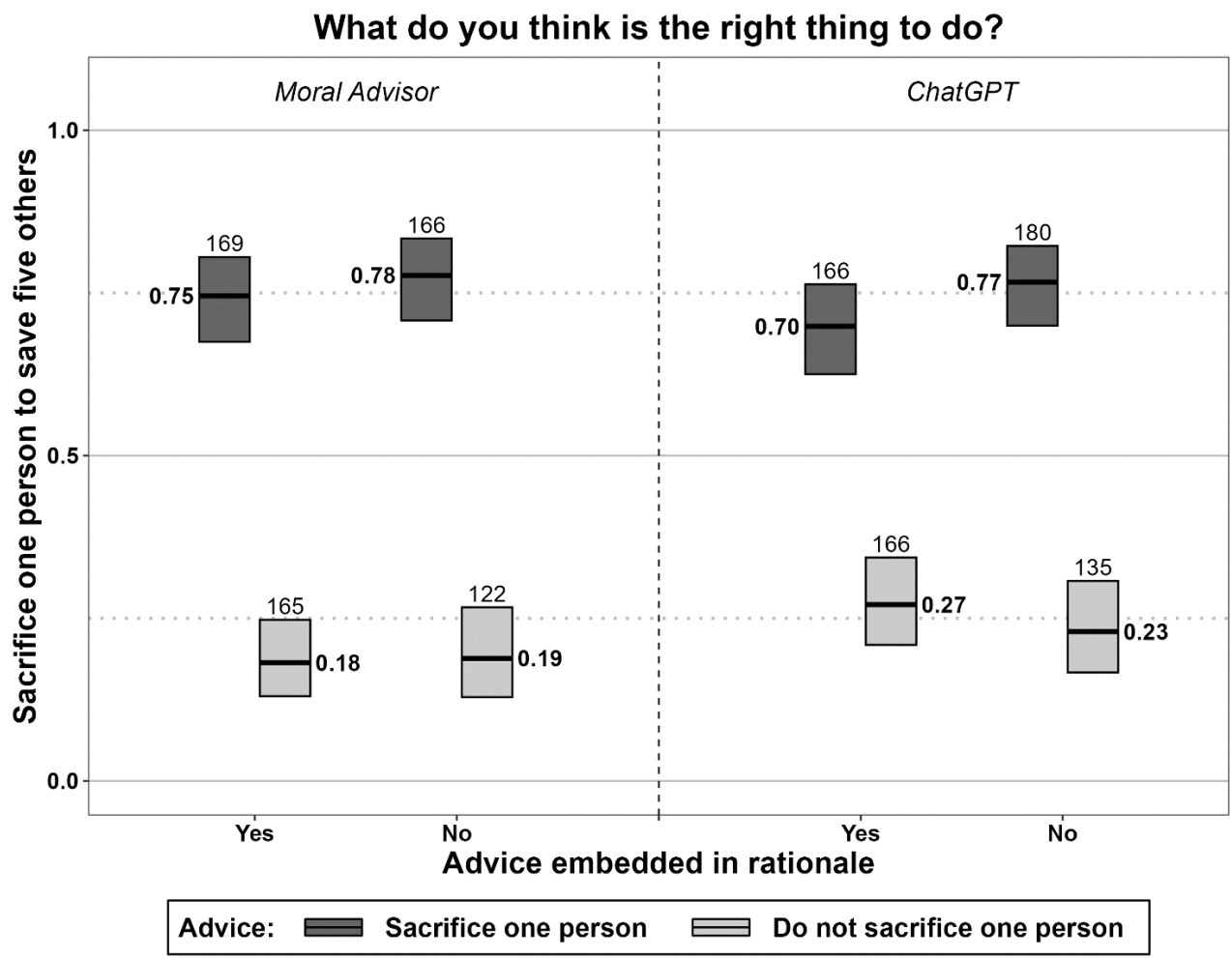

**Figure 2.** Influence of advice on moral judgment. The figure plots the proportions, along with the 95% confidence intervals, of subjects who find sacrificing one person the right thing to do after receiving advice. The numbers of observations figure above the boxes.

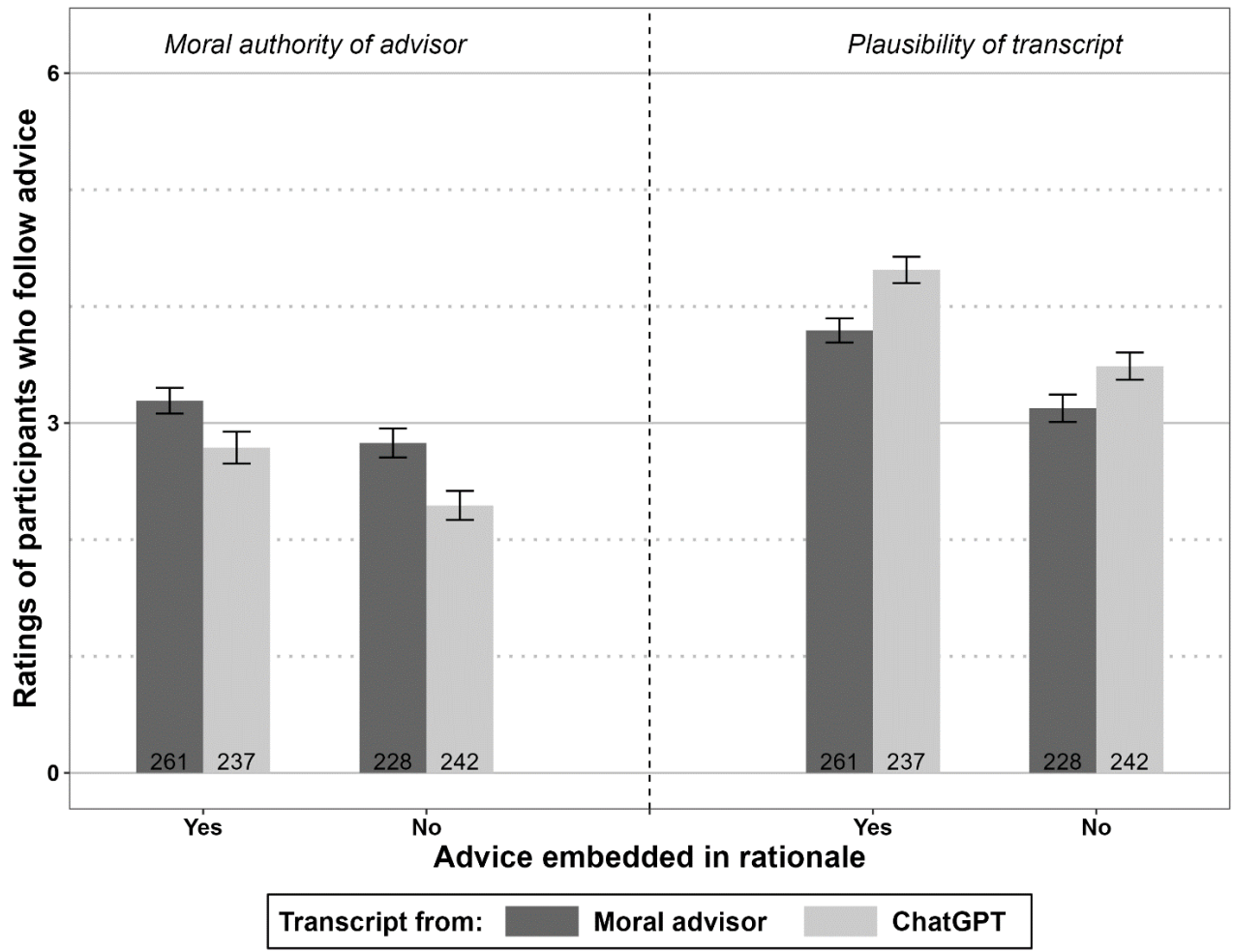

**Figure 3.** Perceived moral authority and plausibility of advice among participants who follow advice. The figure plots the mean ratings and standard errors of the mean as well as the number of subjects at the bottom of each bar.

|  | Moral advisor | | ChatGPT | | |
|  | Argument | | Argument | | |
|  | Yes | No | Yes | No | Statistical testing |
| --- | --- | --- | --- | --- | --- |
| N | 334 | 288 | 332 | 315 |  |
| Age | 45.6 | 45.0 | 43.0 | 44.5 | $F = 1.45, p = 0.23$ |
| Gender |  |  |  |  |  |
| Female | 60% | 57% | 63% | 59% |  |
| Male | 38% | 42% | 35% | 39% | $\chi^2 = 4.52, p = 0.61$ |
| Other/not revealed | 2% | 1% | 2% | 2% |  |
| Ethical | 65.6 | 65.5 | 67.3 | 66.1 | $F = 0.71, p = 0.55$ |

**Table 1.** Summary statistics of information about participants.